\documentclass[pdflatex,sn-mathphys]{sn-jnl}

\usepackage{graphicx}
\usepackage{bm}
\usepackage{bbold}
\usepackage{braket}
\usepackage{amsmath}
\renewcommand\vec{\mathbf} 
\jyear{2022}%

\theoremstyle{thmstyleone}%
\theoremstyle{thmstyletwo}%

\theoremstyle{thmstylethree}%

 \normalbaroutside
\begin{document}

\title[]{Towards Single Atom Computing via High Harmonic Generation}


\author*[1]{\fnm{Gerard} \sur{McCaul}}\email{gmccaul@tulane.edu}
\author[2]{\fnm{Kurt} \sur{Jacobs}}
\author[1]{\fnm{Denys I.} \sur{Bondar}}

\affil*[1]{\orgname{Tulane University}, \orgaddress{\city{Mew Orleans}, \postcode{70118}, \state{Louisiana}, \country{U.S.A.}}}

\affil[2]{\orgdiv{U.S. Army Research Laboratory}, \orgname{Sensors and Electron Devices Directorate}, \orgaddress{ \city{Adelphi}, \postcode{20783}, \state{Maryland}, \country{U.S.A.}}}



\abstract{The development of alternative platforms for computing has been a longstanding goal for physics, and represents a particularly pressing concern as conventional transistors approach the limit of miniaturization. A potential alternative paradigm is that of \emph{reservoir computing}, which leverages unknown, but highly non-linear transformations of input-data to perform computations. This has the advantage that many physical systems exhibit precisely the type of non-linear input-output relationships necessary for them to function as reservoirs. Consequently, the quantum effects which obstruct the further development of silicon electronics become an advantage for a reservoir computer. Here we demonstrate that even the most basic constituents of matter - atoms - can act as a reservoir for computing where all input-output processing is optical, thanks to the phenomenon of High Harmonic Generation (HHG). A prototype single-atom computer for classification problems is proposed, where a classification model is mapped to to an all-optical setup, with linear filters chosen to correspond to the trained model's parameters. We numerically demonstrate that this `all-optical' computer can successfully perform classification tasks, and does so with an accuracy that is strongly dependent on dynamical non-linearities. This may pave the way for the development of petahertz information processing platforms.}

\maketitle
\section*{Introduction}
Since the birth of information theory \cite{Shannon}, there has been an irresistible similarity between computation as a means of processing of information, and the dynamical evolution of physical systems. This isomorphism was striking enough that John Wheeler speculated that the ultimate origin of physical laws may lie in informational dynamics -- that we obtain `it from bit'  \cite{1978}. The cross-pollination of information theory and physics has been particularly fruitful. Considering physical systems as information processors has allowed for the formulation of physical bounds based on informational principles \cite{Lieb1972,LLoyd2014,PRXQuantum.2.010101}, but perhaps the most important consequence of this synthesis has been the realisation that with suitable interpretation, \emph{all} physical systems can act as computers \cite{SethLloyd2000,Lloyd2002}. 

The appeal of using physical systems for computation lies in the trade off it performs between the computational and energetic costs of a problem. An archetypal example of this is the demonstration that an analog `quantum computer' consisting of layers of multi-slit screens can be used to solve NP-complete problems in polynomial time, but with an exponential energetic cost \cite{Aerny1993}. As conventional electronics progress rapidly towards their ultimate limits \cite{HOENEISEN1972819,RADAMSON201819,SIDDIQUI2018107}, the development of such alternative computing platforms is of great importance.

The best known example of a physical computer -- the brain -- has inspired a number of approaches collectively known as neuromorphic computing \cite{Schuman2017}. Within this paradigm, the most relevant framework is that of \emph{reservoir computing} \cite{Lukosevicius2009,Tanaka2019}. A reservoir computer is one in which only the input and output weights connecting to a `reservoir' network are trained. The effect of such a reservoir is to perform a non-linear transformation on the input data, increasing the effective dimensionality of the data. This transformation exploits the crucial property that a set of non-linearly separable data may become linearly separable in a higher dimensional space, facilitating computation \cite{Anco2008, Appeltant2011}.

Given the purpose of a reservoir is to act as a non-linear transformation of inputs, physical systems with a sufficiently non-linear response are natural candidates to fulfil this role  \cite{Jaeger2004}. The existence of such highly non-linear systems has resulted in a number of proposals for physical reservoir computers, harnessing both quantum \cite{Fujii2017, Ghosh2019, Govia2020, Kalfus2021,extra1,extra2,extra3} and classical dynamics \cite{Marcucci2020}. Some of the most promising platforms for reservoir computing are photonic \cite{Pierangeli2021,photoncompute}, as not only have deep learning networks been implemented optically \cite{Shen2017,Opala2019,Hamerly2019}, but some of the core functionalities of an electronic computer, such as hashing \cite{Pappu2002} and programmability \cite{Bogaerts2020}, can be efficiently realised with optics.

 \begin{figure}
\begin{center}
\includegraphics[width=0.9\columnwidth]{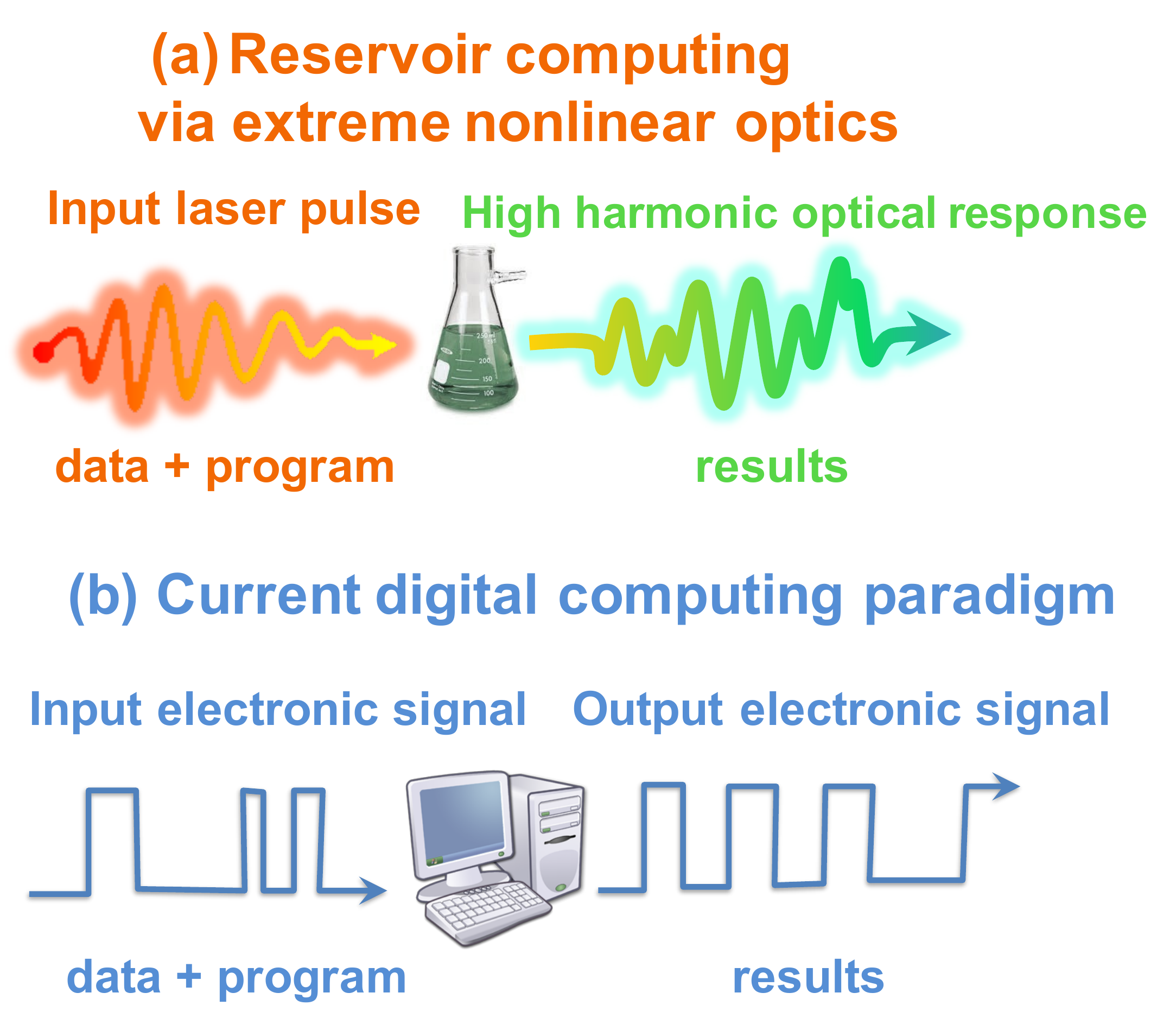}\end{center}\caption{Computations performed via High Harmonic Generation (HHG) are analogous to standard digital methods. HHG is the production of high frequency overtones by driving a system with an intense laser pulse. Figure (a) illustrates the input-output universality of HHG, where a laser pulse (input) can always be found that induces a desired optical response (output) from an arbitrary optically active system. (b)The universality of digital computations implies that all outputs are attainable irrespective of hardware with appropriate software and data.} 
\label{fig:comparison}
\end{figure}

In conjunction with this progress in optical computing, atomic scale control has increased to the point where single atoms can act as transistors \cite{Fuechsle2012}, and be entangled with single photons \cite{Wilk2007}. Given any system with sufficiently non-linear behaviour can serve as a processor for encoded data, this begs the question as to whether it is possible to perform the ultimate act of miniaturisation and implement an optical computer that uses a single atom as its reservoir. In fact, the general public have anticipated such developments with greater foresight than the scientific community. In a recent online educational video concerning atomic networks, one commentator predicted that it would soon be possible to ``run Doom \cite{sanglard2018game,kushner2004masters} on a single atom'' \cite{youtube}.

Computing on a single atom has more than just memetic appeal however -- atomic systems exhibit one of the most extreme non-linear effects yet observed, a phenomenon known as High Harmonic Generation (HHG) \citep{HHGenhance1,HHGenhance2,HHGenhance3,Silva2018}. HHG occurs when a short and intense laser pulse generates responses of up to $\approx 100$ times the frequency of the incident field. It has been shown that both atomic \cite{Campos2017} and solid-state \cite{tracking1,tracking2} systems exhibiting HHG also possess the equally desirable property of \emph{universality}. In a computational context, universality implies that regardless of architecture, any output result can be obtained by suitable input data and software. An analogous physical property is present in HHG systems (see Fig. \ref{fig:comparison}), where it has been shown \cite{Campos2017,tracking1,tracking2,mixing} that regardless of the physical specifics of the system, there always exists an incident laser field (input) that will induce a desired optical response (output). This suggests that a single atom undergoing HHG is a good candidate for reservoir computing.

While the aforementioned reservoir computing schemes all contain physical systems at the heart of their computational network, a similarly physical interpretation of the input and output layers of these networks has often been lacking. This suggests that the use of physical systems as computers requires some attendant electronic computation both to encode information and decode outputs. In other words, a reservoir computer may simply behave like an `ordinary' computer with an intermediate nonlinear element. 

Here we address this, and propose a simple procedure for constructing an `all optical' computation using a single atom exhibiting HHG as the reservoir. Specifically, we demonstrate how one can construct an optical classifier which requires no electronic post-processing to classify data points. Moreover, the proposed setup is similar to those currently employed in HHG experiments \cite{Sommer2016}. 

\section*{Results}
 \begin{figure}
\begin{center}
\includegraphics[width=1.01\columnwidth]{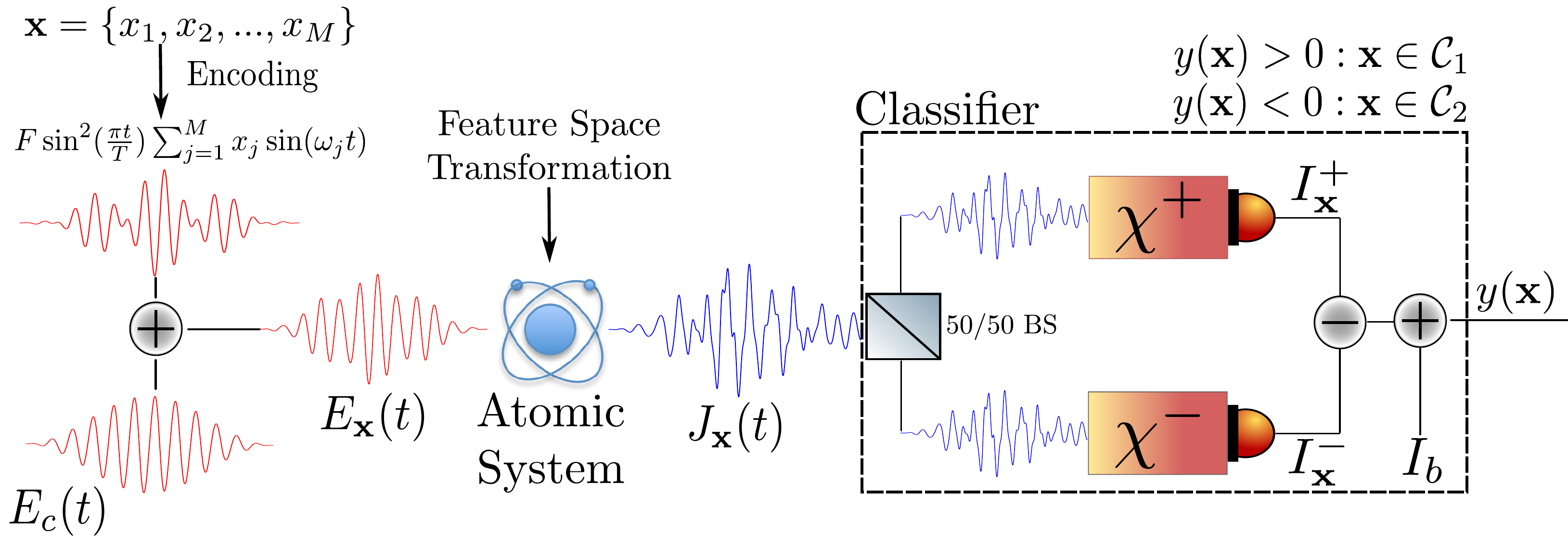}\end{center}\caption{Optical setup for binary classification. After encoding information into the incident laser pulse, the atomic system acts as reservoir, performing a non-linear feature space transformation on the data. Parameter optimisation during training corresponds to the design of linear filters $\chi^\pm$. The difference between the intensity of light passing through these filters (with a constant bias) can then be used to classify the input data.} 
\label{fig:setup}
\end{figure}
 
In order to map a classification problem onto an all-optical setup (see Fig.~\ref{fig:setup}), we first encode the input data $\vec{x}$ into an electromagnetic field to be applied to the atomic system. While this encoding can itself be optimised as a layer in the network, this may result in technically unfeasible pulse shapes.  To ensure experimentally viable laser fields are generated, we therefore prescribe that the data be encoded in a set of enveloped plane waves:
\begin{equation}
\label{eq:encodingpulse}
E_\vec{x}(t)=F\sin^2\left(\frac{\pi t}{T}\right)\left(\sum_{j=1}^M x_j \sin(\omega_j t)\right) +E_c(t).
\end{equation}
Here $T$ is the total duration of the pulse, $F$ is its amplitude and the $\omega_j$ are a set of $M$ frequencies matching the dimensionality of the data. An additional \emph{carrier field}
\begin{align}
E_c(t)=F_c\sin^2\left(\frac{\pi t}{T}\right)\sin(\omega_c t)
\end{align}
is also included, so as to guarantee the full field will induce a non-linear response in the atomic system \cite{Marcucci2020}.

In general the precise physical system used is immaterial, provided that it exhibits a sufficiently non-linear response to act as a reservoir in the computation \cite{Jaeger2004}. For specificity, we use here a single active electron approximation, such that the atom's dynamics can be represented as a single particle moving in an effective potential \cite{Bauer1997}
\begin{equation}
    V(\hat{r})=-\frac{Z_e}{\sqrt{\hat{r}^2 + a^2}},
\end{equation}
where we have employed $\hat{r}$ as the 1D position operator to distinguish it from the input data vector $\vec{x}$. For the sake of specificity we take $Z_e=1$ and $a^2=2.37$ a.u., a parametrisation corresponding to an argon atom. Finally, the field is coupled to the system via the dipole approximation, such that the total system Hamiltonian is described by:
\begin{equation}
    \hat{H}(t)=\frac{1}{2}\hat{p}^2 +V(\hat{r}) -\hat{r}E_{\vec{x}}(t).
\end{equation} 

The system simulation is initiated in the ground state, and for each data point is evolved using the Hamiltonian with the encoded pulse $E_{\vec{x}}(t)$. The optical response is then calculated, using $J(\omega)=\mathcal{F}_{t\to\omega}\left[\frac{{\rm d}}{{\rm d} t} \left<p\right>\right]$, where $\mathcal{F}_{t\to\omega}$ denotes the Fourier transform. Using this response, a set of optical elements can then be constructed such that the the input data is classified purely from a set of measured intensities.

To show this, let us first briefly examine the mechanics of a maximum margin classifier. The extension to more sophisticated classifiers (such as a support vector machine with slack variables) will not materially affect the mapping between the classifier and physical quantities \cite{bishop2006pattern}. Furthermore, we will focus exclusively on mapping a binary classification problem, as any multiclass classifier may be constructed from a set of binary classifications \cite{cristianini2000an}. Consider first a model mapping an input $\vec{x}$ to an output $y(\vec{x})$: 
\begin{equation}
\label{eq:linearmodel}
y(\vec{x})=\vec{w}^T\phi(\vec{x}) + b
\end{equation}
where $\phi(\vec{x})$ is a data point transformed to some feature space, and $b$ is a bias parameter. We wish to find the weights $\vec{w}$ such that if the data point  $\vec{x}_n$ in the training set of $N$ data points is classified by a target value $t_n=1$ then $y(\vec{x}_n) >0$, and conversely if $t_n=-1$ then $y(\vec{x}_n) < 0$. These weights are found by maximising the margin between the $y=0$ hyperplane and the nearest data points. This problem is equivalent to minimising $\left|\vec{w}\right|^2$, subject to the inequality constraint
\begin{equation}
    t_n\left(\vec{w}^T\phi(\vec{x}_n)+b\right)\geq 1, \ \ n=1,..,N.
\end{equation}

Such a constrained minimisation problem can be solved with the introduction of Lagrange multipliers $a_n$. After optimisation of the $a_n$, we can re-express Eq.~\eqref{eq:linearmodel} as:
\begin{equation}
\label{eq:kernel_classifier}
   y(\vec{x})=\sum_{n=1}^N a_n t_n \phi^T(\vec{x}_n)\phi(\vec{x}) + b. 
\end{equation}
where the parameters $a_n$ parameters obey the Karush-Kuhn-Tucker conditions \cite{kuhn1951nonlinear,avriel2003nonlinear}:
\begin{align}
    a_n & \geq 0, \\
    t_n y(\vec{x}_n) -1  & \geq 0, \\
    a_n \left( t_n y(\vec{x}_n) -1\right)  & = 0.
\end{align}
The last condition is of particular importance, as it results in $a_n=0$ for all points except those lying closest to the $y(\vec{x})$ hyperplane \cite{bishop2006pattern}.

Anticipating the mapping of Eq.~\eqref{eq:kernel_classifier} to an optical process, we define two sets whose union contains all the non-zero $a_n$ -- $\mathcal{S}^+$ for those points where $t_n=+1$, and $\mathcal{S}^-$ for points corresponding to $t_n=-1$. Finally, we make the mappings: 
\begin{align}
b&\to I_b, \\
    \phi(\vec{x}) &\to \left|J_\vec{x}(\omega)\right|^2, \\ 
     \label{eq:filtermapping}\sum_{a_n\in \mathcal{S^{\pm}}} a_n \phi^T(\vec{x}_n) &\to \int d\omega \left|\chi^{\pm}(\omega)\right|^2,
\end{align}
where $\chi^\pm (\omega)$ describes the susceptibility of a linear filter \cite{boyd_nonlinear_2008}. The intensity $I$ of light after the response $J(\omega)$ is passed through such a filter is given by:
\begin{equation}
    I_\vec{x}^\pm=\int{\rm d}\omega \ \left|\chi^{\pm}(\omega)\right|^2 \left|J_\vec{x}(\omega)\right|^2.
\end{equation}
 meaning that Eq.~\eqref{eq:kernel_classifier} can be remapped to
 \begin{equation}
     y(\vec{x}) =I_\vec{x}^+ -I_\vec{x}^- +I_b.
 \end{equation}
Consequently, the classification problem can be expressed in optical terms by splitting the response $J(\omega)$ and measuring the intensities after passing it separately through the filters $\chi^+$ and $\chi^-$. The data point can then be classified according to whether the difference between these intensities is greater than $I_b$. 

Finding both $I_b$ and $\chi^\pm$ first requires an electronic computation, taking a set of training data and classifying the simulated set of $\left\{J_\vec{x}(\omega)\right\}$ using a quadratic kernel. Once the optimal values for $a_n$ have been found, this can be translated directly into both a value for $I_b$ and a susceptibility profile for the filters. Using this information, new data may be encoded and classified using the single-atom computer setup shown in Fig. \ref{fig:setup}. Given the (almost) zero computational overhead involved, new data may be classified at close to light speed.  

We now detail the application of this method to several classification problems, simulating the atomic system's optical response and training on this data to generate susceptibility profiles. In all cases, data is normalised in order to ensure $F\sum^M_{j=1} \left|x_j\right| < F_c$, so that any one dimension of the data encoded in the pulse does not dominate over the carrier wave. We choose a carrier wave with $\omega_c = 0.06$ a.u., and encoding frequencies are linearly spaced between $\omega_c$ and the largest encoding frequency $\omega_M$.

Additionally, in order to ensure the atomic system is acting as a reservoir, and that the classifier is being trained on a non-linear function of the input data, all sub-linear frequencies (in relation to the encoding frequencies) are pre-filtered from the response before training: $\bar{J}(\omega)= J(\omega)\Theta(\omega-\omega_M)$, where $\Theta(z)$ is the Heaviside function. Finally, the accuracy of the trained models is assessed using their f1-score, the harmonic mean of the trained model's precision and recall on test data \cite{derczynski-2016-complementarity}. An f1-score of 1 indicates perfect classification of data, while a score of 0 corresponds to perfect misclassification. 

\subsection*{Digit Recognition}
As a first test of the method, we use the archetypal dataset of handwritten digits, each described by a 64 dimensional vector \cite{Dua:2019}. While the optical system is quite capable of classifying each digit, for the sake of clarity and simplicity we reduce the problem to binary classification. To do so, we assign any digit $< 5$ to class $\mathcal{C}_1$, and all others to $\mathcal{C}_2$, i.e. $\{0,1,2,3,4\} = \mathcal{C}_1, \ \ \{5,6,7,8,9\} = \mathcal{C}_2$.

The dataset contains $N=1797$ points, and unless otherwise stated, 10\% of the data is used for training and 90\% for testing. Each point is encoded according to Eq.~\eqref{eq:encodingpulse}, generating a simulated response $\left|J_\vec{x}(\omega)\right|^2$ used to both train and test the model. In all cases the bottom encoding frequency is taken to be $\omega_1=2\omega_c$. 

 \begin{figure}
\begin{center}
\includegraphics[width=1\columnwidth]{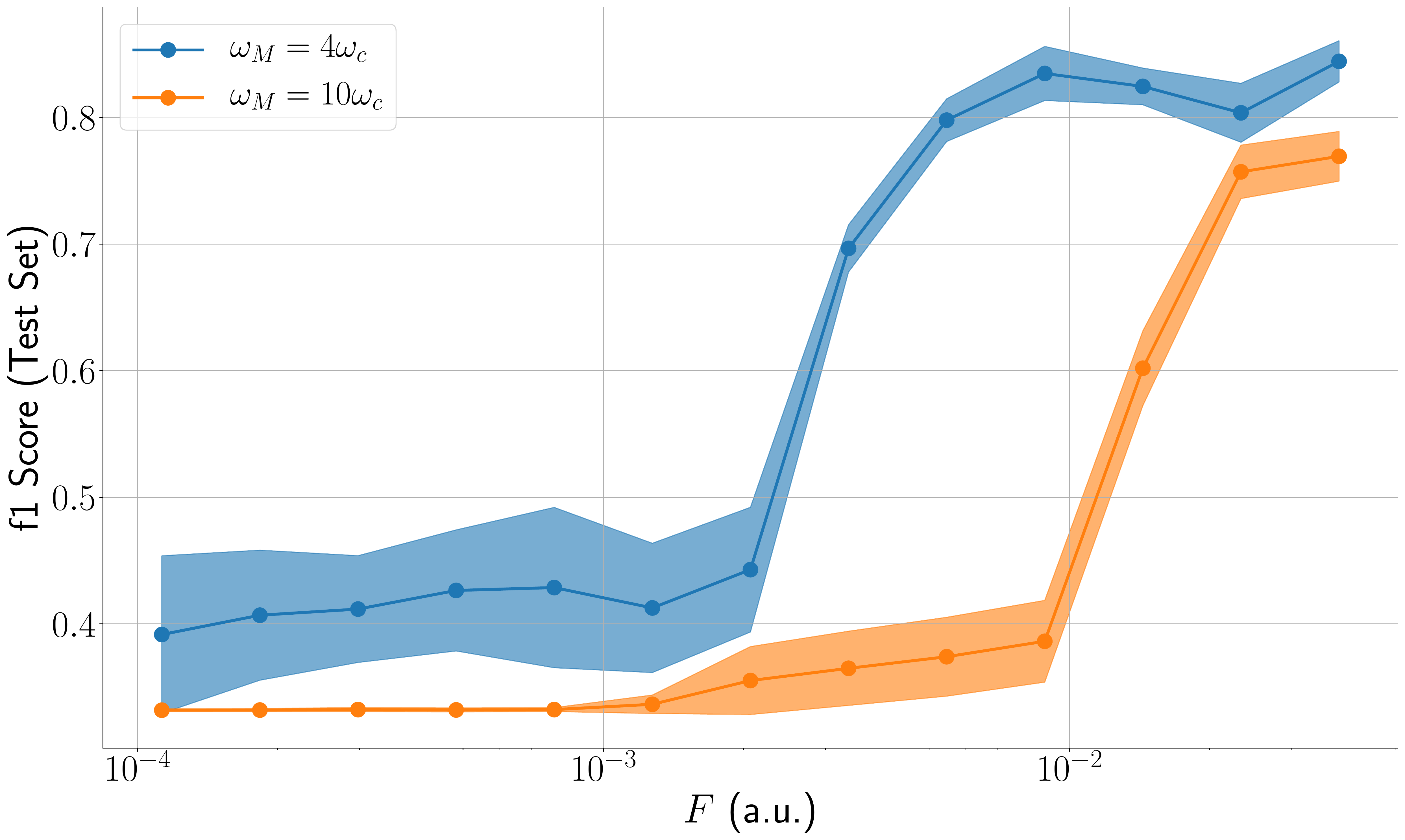}\end{center}\caption{Example of f1-score dependence on field amplitude $F$ for binary classification of digits. Averages are calculated over twenty independent runs, with the standard deviation highlighted by the shaded areas. The ability of the model to successfully classify points is heavily dependent on whether a threshold amplitude for the input pulse has been reached. This threshold is itself dependent on $\omega_M$, the highest frequency used to encode data.} 
\label{fig:fscordigits}
\end{figure}

\begin{figure}
\begin{center}
\includegraphics[width=1\columnwidth]{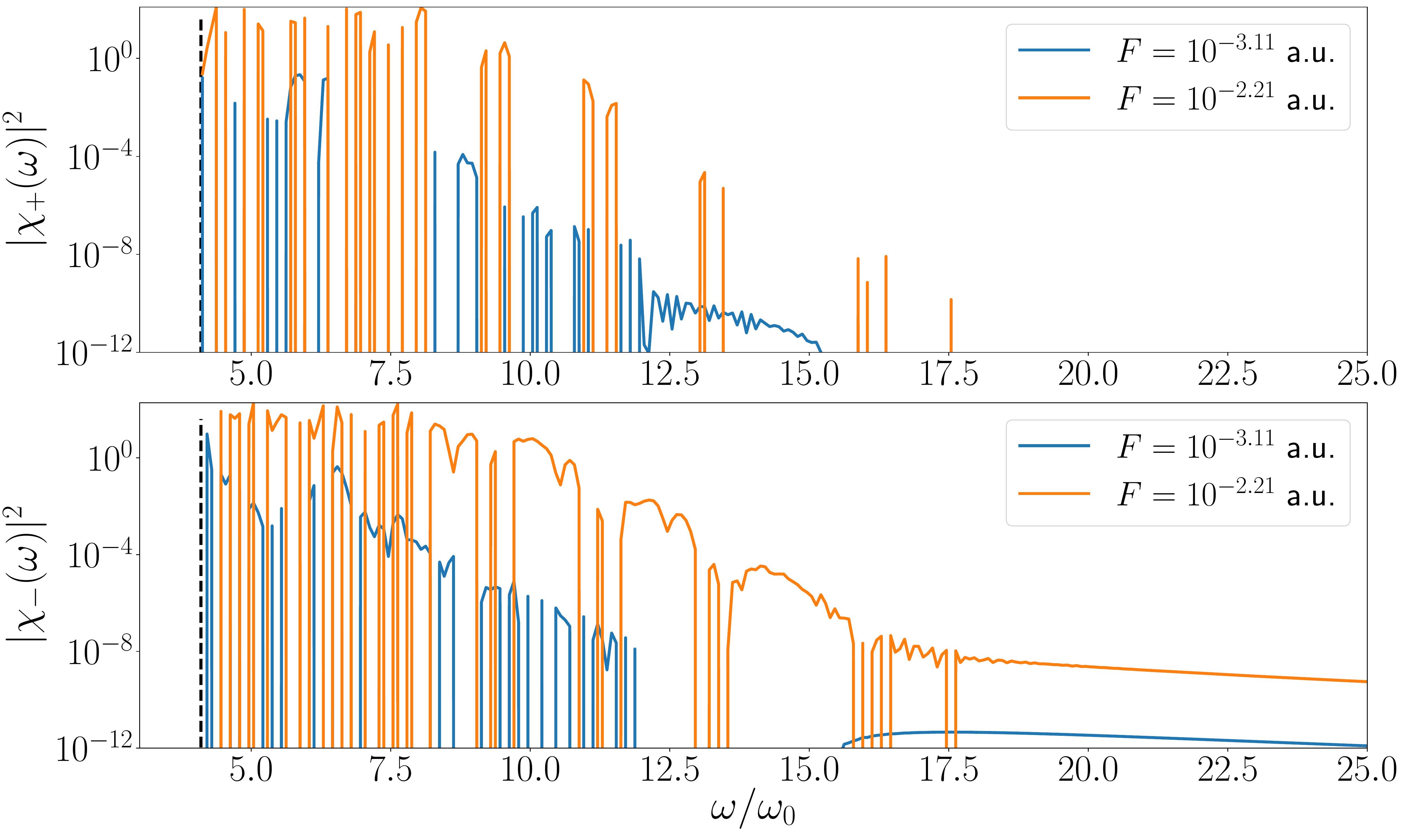}\end{center}\caption{Example susceptibilities for the encoding $\omega_M = 4\omega_c$ for digit recognition. These profiles are constructed from the weights of the optimised model, which in turn depend on the encoding field amplitude $F$. The dashed black-line indicates $\omega_M$. } 
\label{fig:profilesdigits}
\end{figure}

Figure ~\ref{fig:fscordigits} shows the trained model's accuracy as a function of the laser field amplitude, demonstrating the classifier's success is heavily dependent on the incident light's intensity. Given all sub-linear frequencies are filtered out, the f1-score serves as an indicator of the degree of non-linearity present in the response. In order to accurately classify the data, the field amplitude must be sufficiently large that the information in the lowest encoding frequency $\omega_1$ is present in the response frequencies $\omega > \omega_M$. For this reason, while a larger $\omega_M$ makes the different encoding frequencies more distinguishable, it also requires a more intense laser to ensure all information is transmitted to the high harmonics. Once the model has been trained, it is possible to extract susceptibility profiles for the filters via Eq.~\eqref{eq:filtermapping}, as shown in Fig.~\ref{fig:profilesdigits}. As the classifier accuracy is dependent on $F$, it is not surprising that the trained model weights (and hence $\chi$) will change with the field amplitude.  

\subsection*{Iris Classification}
Here we use the atomic computer for a ternary classification problem. Specifically, we use the iris dataset \cite{duda1973pattern}, where each data point describes four measurements of one of three types of iris flower. This dataset has the advantage that only one of the three classes is linearly separable from the other two, and thus the atomic system must truly act as a reservoir in order to sucessfully classify points. The dataset consists of 150 data points (50 of each type of flower), with the same training fraction and bottom encoding frequency as for digit recognition.   

 \begin{figure}
\begin{center}
\includegraphics[width=1\columnwidth]{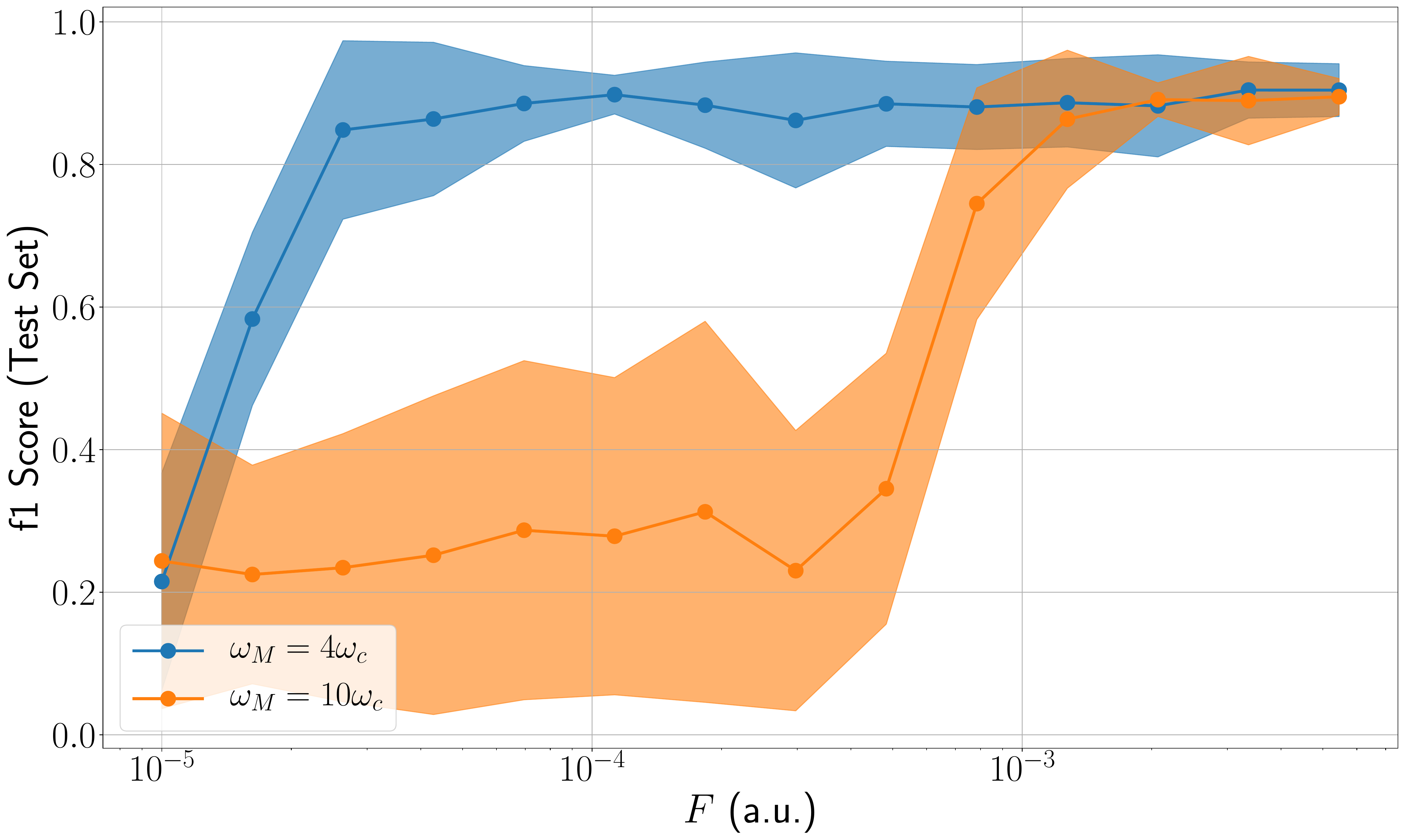}\end{center}\caption{Example of average f1-score dependence on field amplitude $F$ for ternary classification of iris flowers. Averages are calculated over twenty independent runs, with the standard deviation highlighted by the shaded areas.  Once again, for a larger top encoding frequency $\omega_M=10\omega_c$, a threshold laser amplitude must be reached before the data can be accurately classified} 
\label{fig:fscoreiris}
\end{figure}

\begin{figure}
\begin{center}
\includegraphics[width=1\columnwidth]{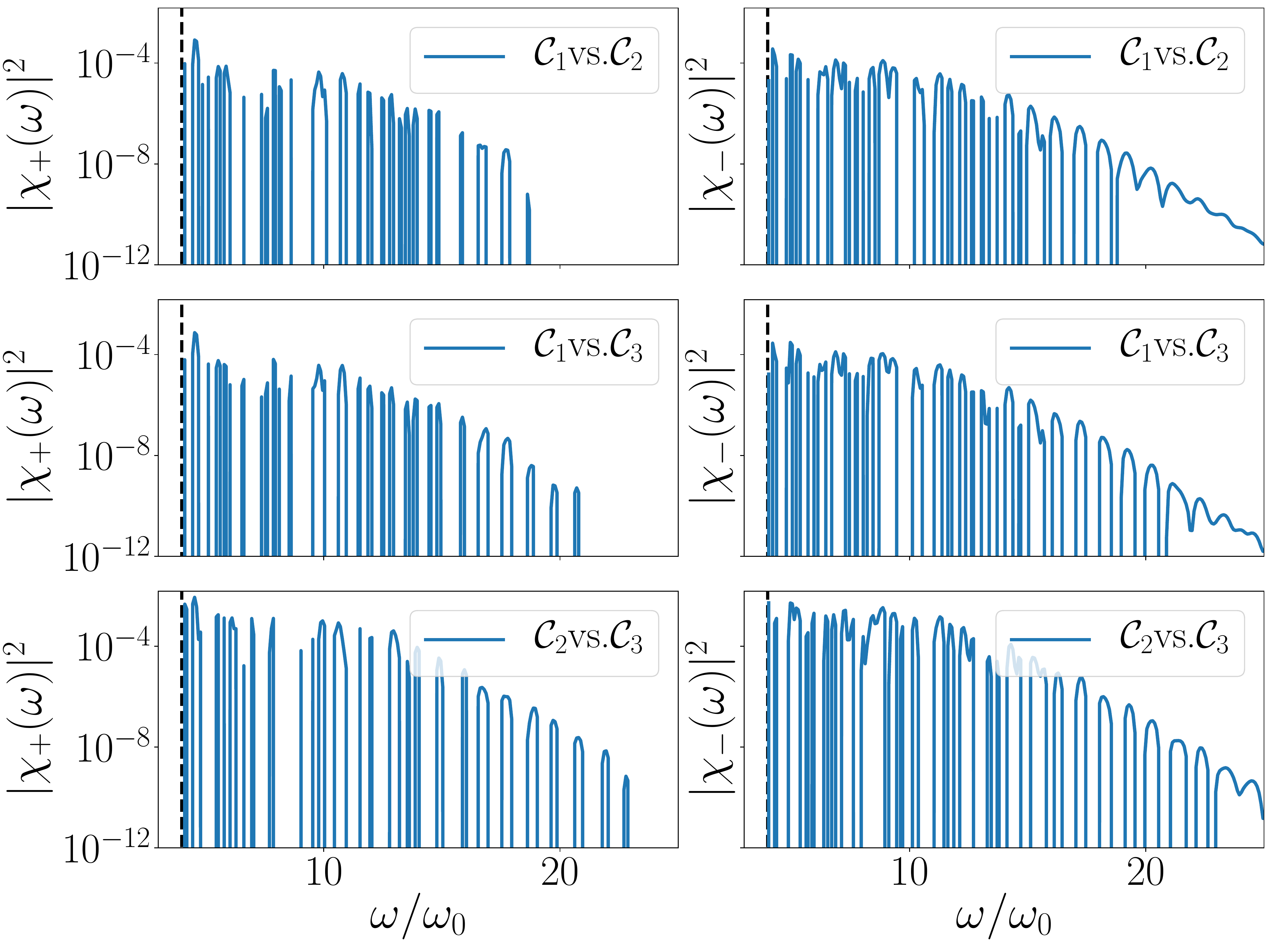}\end{center}\caption{Example susceptibilities for the encoding $\omega_M = 4\omega_c$ for digit recognition. These profiles are constructed from the weights of the optimised model, which in turn depend on the encoding field amplitude $F$. The dashed black-line indicates $\omega_M$. } 
\label{fig:profilesiris}
\end{figure}

Figure~\ref{fig:fscoreiris} once again demonstrates the dependence of the model accuracy on both $F$ and $\omega_M$. While $\omega_M=4\omega_c$ produces good results across the range of intensities considered, the $\omega_M=10\omega_c$ has a clear threshold amplitude, below which the model cannot make accurate predictions. As a multiclass problem, the optical computer requires three sets of $\chi_\pm$ filters. This is due to ternary classification being equivalent to a set of three `one versus one' binary classifications. An example set of these filters is shown in Fig.~\ref{fig:profilesiris}, and are once again dependent on $F$ (not shown).  

\subsection*{Energetic efficiency} 
{
Here we consider the energetic cost of a computation on a single atom as compared to the equivalent procedure on standard architecture. Assessing this is not entirely straightforward, given that the model employed uses an effective description of a classical field. Despite this, it is at least possible to calculate the efficiency of the computation relative to the input pulse. For each data point processed, we consider three quantities - the change in energy of the atomic system over its evolution $\Delta E$, the proportion of the output pulse contained in the fully nonlinear part of the spectrum $Q_\vec{x}$:
\begin{equation}
    Q_\vec{x}=\frac{\int^\infty_{\omega_M} {\rm d}\omega \  \left|J_\vec{x}(\omega)\right|^2}{\int_0^{\omega_M} {\rm d}\omega \  \left|J_\vec{x}(\omega)\right|^2},
\end{equation}
and $R$, the proportion of the non-linear response $\bar{J}(\omega)$ that is transmitted through the classification filters (whose maximum is normalised to 1):
\begin{equation}
    R_\vec{x}= \frac{\int^\infty_{\omega_M} {\rm d}\omega \ \sum_j^N \sum_{\sigma}\left|\chi_j^{\sigma}(\omega)\right|^2 \left|J_\vec{x}(\omega)\right|^2}{2N\int^\infty_{\omega_M} {\rm d}\omega \left|J_\vec{x}(\omega)\right|^2}
\end{equation}
where $j$ labels each binary classifier filter $\chi^\sigma_j$ required.} 

\begin{figure}
\begin{center}
\includegraphics[width=1\columnwidth]{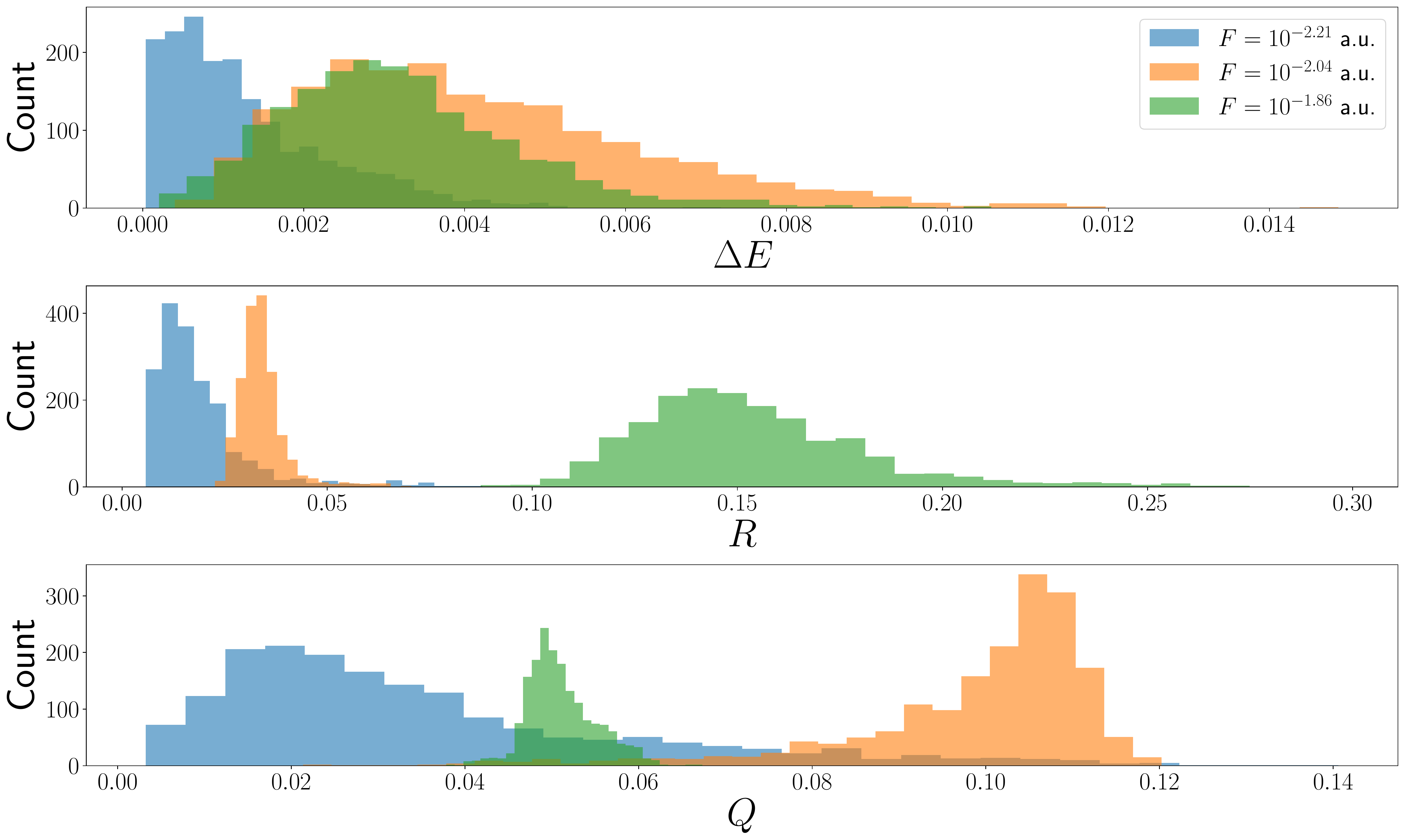}\end{center}\caption{Distributions of $\Delta E$, $R$ and $Q$ for the digit dataset using three example input amplitudes $F$. Here $\omega_M=4\omega_c$} 
\label{fig:histograms}
\end{figure}

{Here, $\Delta E$ allows one to estimate the number of laser photons absorbed by the atomic system, and hence the degree of the input laser's energy that is contained in the output response. The product of $Q_\vec{x} R_\vec{x}$ can then be used to determine what proportion of this energy is usefully used in the classification process. These figures of merit for digit classification are shown in Fig. \ref{fig:histograms}. Considering the distribution of $\Delta E$, we find that across all data points, the energy absorbed by the system is less than that of the carrier frequency $\omega_c$, regardless of the input amplitudes $F$. This implies that the total energy absorbed by the atom is negligible compared to the laser's total energy, and hence the output response will have an almost identical energy to the input pulse.}

{It is worth noting that while one expects the degree of HHG to increase with $F$, this relationship will also depend on the input frequency. Given the encoded pulse consists of multiple frequencies however, a proportion of the non-linear response due to the lower encoding frequencies will still lie below $\omega_M$. This means that while in a monochromatic pulse $Q$ would increase with $F$, here we find that at the largest values of $F$ considered, the average $Q$ decreases again.} 

{Finally, combining $R$ and $Q$ we find that only $\sim$1\% (on average) of the total energy injected into the system is ultimately used for classification. This suggests that the vast majority of the light generated by the input pulse is free to be used for other purposes, potentially yielding much greater overall efficiencies. Given the relative inefficiency and intense energy consumption of high powered lasers however, it is clear that even if most of the energy of the output pulse can be recycled, its efficiency will still fall well below that of conventional computing.}

\section*{Discussion \label{sec:Discussion}}
Conventional computing is rapidly approaching the point at which tunnelling effects render further miniaturisation of transistors impossible. The development of alternative computing platforms is therefore of urgent concern. In this manuscript we have outlined a potential alternative architecture, which seeks to utilise rather than avoid quantum dynamics. In particular, a mapping was established between optimised network weights and optical elements. Consequently, we have demonstrated the feasibility of constructing an `all optical' single-atom computer. 

This approach was tested with several benchmark classification problems, where in each case the ability of the atomic computer to successfully classify data was strongly dependent on both the top encoding frequency $\omega_M$ and the laser amplitude $F$. After training the model on the simulated responses $\left|J_\vec{x}(\omega)\right|^2$, it is possible to extract susceptibility profiles for linear filters via the optimised weights of the trained model. These filters can then be used to perform classification of encoded data solely via an intensity measurement of the optical response after passing through the filters. 

The potential benefits of such a computing system are numerous. Atomic systems exhibiting HHG not only possess an extreme non-linearity that makes them ideal reservoirs, but display input-output universality. This guarantees that it is in-principle always possible to encode a computational problem into an optical setup of the type detailed here. This has the distinct advantage that computations are performed at the speed of the atomic dynamics. Consequently, the ultrafast  dynamics of HHG raises the possibility of processing information at petahertz speeds \cite{Li2020,Corkum2007,RevModPhys.81.163,Sommer2016}. 

{One drawback of considering a single atom is that it is likely to produce too few high harmonic photons to be experimentally detectable. This scheme can however be extended to the use of a much larger ensemble of atoms as a reservoir. While any theoretical calculation would be greatly complicated, the training process for filter design can be carried out entirely experimentally, using the obtained output spectra from the training set to fit the classification model's parameters (and hence the filters).}

{As stated in the introduction, this ultrafast processing must be traded off for the energetic efficiency. Lasers of sufficient intensity to generate non-linear responses are themselves not efficient, and the energy consumed processing a data point in this framework will not yet compare favourably to that used by conventional computing. There is some likelihood however that this will change in future, given that the development of intense broadband laser sources are being actively developed \cite{elu2020seven}. Furthermore, while here the linear part of the response is neglected, this can instead be utilised. Recent work has demonstrated that cavity assisted HHG has the potential to dramatically improve the efficiency of the process \cite{Hogner:19}. In this context, future investigations may focus on whether the neglected linear part of the response can be used as part of an optical cavity, and greatly improve the efficiency of all optical calculations.}
\begin{figure}
\begin{center}
\includegraphics[width=1\columnwidth]{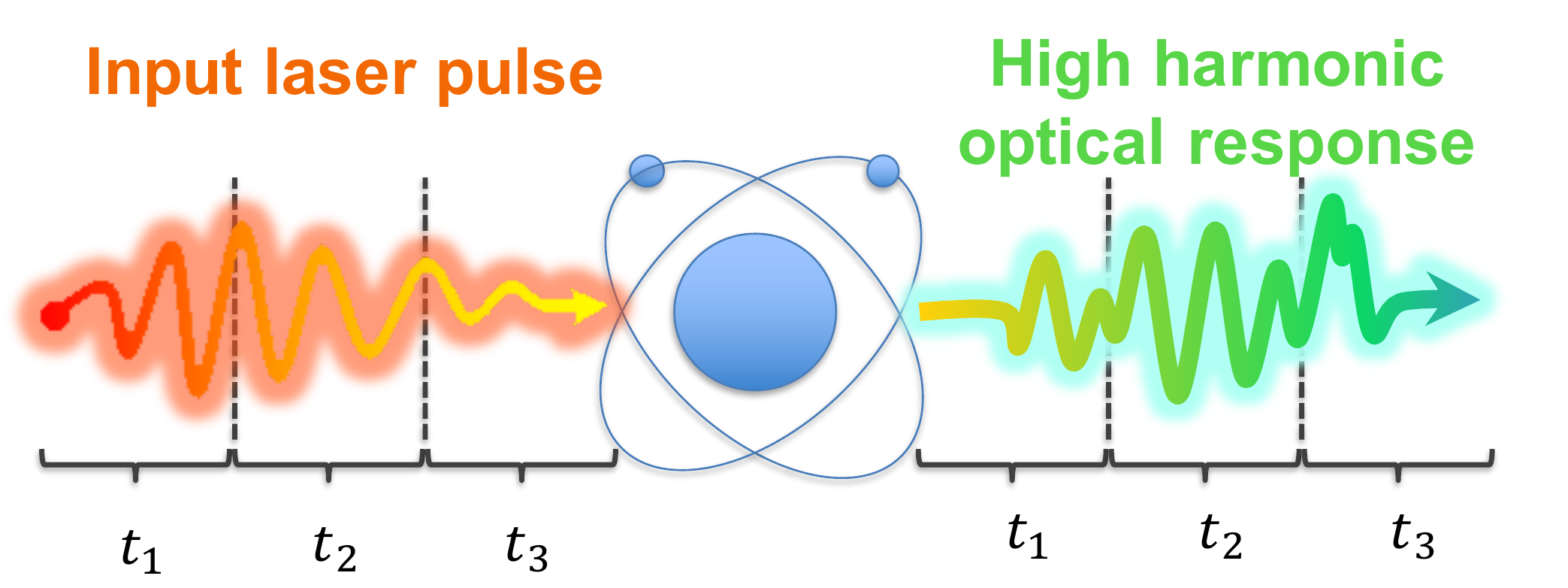}\end{center}
\caption{Connection between a single atom computer and a quantum computer. Here both input and output our divided into three time periods corresponding to state preparation, emulation of a unitary gate, and finally tomography to readout the result of the computation.} 
\label{fig:ConnectionQC}
\end{figure}

{Finally, we would like to conjecture an equivalence between a quantum computer and a single atom computer via HHG. To model the latter, classical electromagnetic radiation has been employed for both input and output. Only the coupling between light and matter has been described quantum mechanically. Thus, quantum effects serve as a source of extreme optical nonlinearities. We conjecture that quantum information processing can be performed on such a platform if we slice the input pulse as well as output into three time intervals $t_1$, $t_2$, and $t_3$ as shown in  Fig.~\ref{fig:ConnectionQC}: Time slot $t_1$ contains a shaped laser pulse that prepares the initial state necessary for a quantum computation; $t_2$ -- a shaped pulse generating a desired unitary gate. Both the pulses can be obtained, e.g., by using optimal quantum control \cite{d2007introduction}. The remaining time interval $t_3$ is used to perform tomography of the final quantum state \cite{yang_complete_2020}, which contains a result of quantum computation. This analogy between quantum and single atom computer offers a natural way for implementing a hybrid computation.}

\section*{Acknowledgement}
The authors would like to thank Claudio Conti for helpful discussions. G.M. would also like to thank Josh McNamee for their suggestion of Ref. \cite{sanglard2018game}. This work has been generously supported by Army Research Office (ARO) (grant W911NF-19-1-0377, program manager Dr.~James Joseph, and grant W911NF-21-2-0139). The views and conclusions contained in this document are those of the authors and should not be interpreted as representing the official policies, either expressed or implied, of ARO or the U.S. Government. The U.S. Government is authorized to reproduce and distribute reprints for Government purposes notwithstanding any copyright notation herein.

\section*{Data Availability}
Both the data and code that support the findings of this study are available from the corresponding author upon reasonable request.


\providecommand{\noopsort}[1]{}\providecommand{\singleletter}[1]{#1}%

\end{document}